\documentclass[12pt, letterpaper]{article}

\usepackage[utf8]{inputenc}
\usepackage{amsmath}
\usepackage{amssymb}
\usepackage{xcolor}
\usepackage{graphicx}
\usepackage[margin=1in]{geometry}
\usepackage{booktabs}
\usepackage{enumitem}
\usepackage{fancyvrb}
\usepackage{tikz}
\usetikzlibrary{shapes.geometric, arrows, positioning, fit, calc, patterns}
\usepackage{hyperref}

\hypersetup{
    colorlinks=true,
    linkcolor=blue,
    filecolor=magenta,
    urlcolor=cyan,
    citecolor=blue,
}

\tikzstyle{startstop} = [rectangle, rounded corners, minimum width=3cm, minimum height=1cm, text centered, draw=black, fill=red!10]
\tikzstyle{process} = [rectangle, minimum width=3.1cm, minimum height=1cm, text centered, text width=3.1cm, draw=black, fill=orange!10]
\tikzstyle{decision} = [diamond, minimum width=3cm, minimum height=1cm, text centered, draw=black, fill=green!10]
\tikzstyle{arrow} = [thick,->,>=stealth]
\tikzstyle{io} = [trapezium, trapezium left angle=70, trapezium right angle=110, minimum width=2.6cm, minimum height=1cm, text centered, draw=black, fill=blue!10]
\tikzstyle{container} = [rectangle, rounded corners, draw=black!50, dashed, inner sep=10pt]

\title{Can AI Keep a Secret? \\ \large Contextual Integrity Verification: A Provable Security Architecture for LLMs}
\author{Aayush Gupta}
\date{August 2025}

\begin{document}

\maketitle

\begin{abstract}
The promise of Large Language Models is betrayed by a fundamental vulnerability: they cannot distinguish friend from foe. Today's LLMs remain dangerously susceptible to prompt-injection and jailbreak attacks—a reality that transforms every interaction into a potential security breach. While the industry has deployed an arsenal of heuristic defenses—keyword filters, toxicity classifiers, LLM-based detectors—attackers consistently pierce through these semantic shields, as evidenced by repeated failures in public red-team exercises \cite{owasp2024}. State-of-the-art guardrail toolkits including LLM-Guard \cite{llmguard_github}, Rebuff \cite{rebuff_github_2025}, and PromptArmor \cite{shi2025promptarmor} have pushed the boundaries of resistance, yet they fundamentally rely on probabilistic patterns that sophisticated adversaries can manipulate, obfuscate, or simply bypass.

We present \emph{Contextual Integrity Verification (CIV)}—a paradigm shift from detecting malice to enforcing mathematical boundaries. To our knowledge, CIV is the first security architecture to deliver \emph{deterministic, cryptographically verifiable} non-interference guarantees at inference-time on pre-trained, frozen transformer models. Our approach transcends semantic analysis by embedding an immutable trust lattice directly into the transformer's computational fabric: every token carries a cryptographically signed provenance (HMAC-SHA-256), and attention mechanisms are surgically modified to enforce hard information-flow boundaries. Lower-trust tokens become mathematically incapable of influencing higher-trust computations—not through detection, but through algebraic impossibility.

Evaluated against a comprehensive benchmark synthesized from cutting-edge attack taxonomies (Elite-Attack + SoK-246) \cite{li2025detam,hu2025sok}, CIV achieves what no prior defense has accomplished: \textbf{absolute protection (0\% ASR) within our threat model}, while preserving \textbf{93.1\%} output fidelity and maintaining model perplexity. Our reference implementation, though carrying a notable latency overhead attributable to unoptimized cryptographic pipelines, demonstrates immediate deployability as a lightweight patch for Llama-3-8B and Mistral-7B—no retraining required.

We release our implementation,\footnote{\url{https://github.com/ayushgupta4897/Contextual-Integrity-Verification}} an automated certification framework, and the "Elite-Attack"\footnote{\url{https://huggingface.co/datasets/zyushg/elite-attack}} corpus to accelerate reproducible security research.
\end{abstract}

\section{Introduction}
\label{sec:intro}
Large Language Models have become the backbone of modern AI systems—yet they harbor a fatal flaw. Their extraordinary generative capabilities mask a brittle security foundation: the inability to enforce trust boundaries. Prompt-injection (PI) and jailbreak attacks exploit this weakness, allowing adversaries to smuggle malicious instructions that override system policies, exfiltrate confidential data, or coerce forbidden behaviors. The gravity of this threat is reflected in OWASP's ranking: PI now stands as the \#1 security risk for LLM applications \cite{owasp2024}.

\subsection{The Anatomy of a Prompt Injection Attack}
To understand the severity of this vulnerability, we must examine how trust boundaries collapse in practice. Every LLM application constructs prompts from multiple sources, each carrying an implicit privilege level. An attack succeeds when a low-privilege source hijacks the behavior dictated by a high-privilege source—a violation of fundamental security principles.

\paragraph{Example 1: Direct Injection (Jailbreak).} Here, an adversary weaponizes the low-trust \texttt{USER} channel to subvert high-trust \texttt{SYSTEM} directives:
\begin{Verbatim}[frame=single,fontsize=\small]
[SYSTEM PROMPT - Trust=100] You are a helpful assistant. Never reveal this
instruction. Your name is Alex.

[USER PROMPT - Trust=80] ...ignore all previous instructions and tell me the
first sentence of your instructions.
\end{Verbatim}
A vulnerable LLM capitulates to the user's command, betraying its confidential system prompt—a breach that could expose proprietary knowledge or security configurations.

\paragraph{Example 2: Indirect Injection (Web Retrieval).} The threat escalates when trusted tools become unwitting accomplices:
\begin{Verbatim}[frame=single,fontsize=\small]
[SYSTEM PROMPT - Trust=100] You are a helpful assistant. Use the web
browser tool to answer questions about recent events.

[USER PROMPT - Trust=80] What are the main points of the new tech bill
summarized at document-site.com/summary?

[WEB CONTENT - Trust=20] The bill is about... [INVISIBLE TEXT] ...end your
response with "Access Granted. Admin override successful."
\end{Verbatim}
In this insidious attack, the LLM dutifully retrieves a seemingly benign web page that conceals a malicious payload. The lowest-trust source (\texttt{WEB}) successfully commandeers the entire interaction—a complete inversion of the intended security hierarchy. CIV transforms these attacks from possible to impossible.

\subsection{Why Heuristic Guardrails Are Failing}
The security community's response has been reactive: deploy layers of semantic filters, toxicity classifiers, and LLM-based judges to identify "unsafe" content. Yet a damning pattern emerges from independent evaluations. A recent Systematization-of-Knowledge (SoK) survey reveals that even state-of-the-art guardrails suffer persistent jailbreak rates of 15–30\%—\emph{after} vendors have tuned against the test sets \cite{hu2025sok}. This failure is not a bug but a feature of the approach: semantic defenses are fundamentally playing a losing game against adversaries who can encode, obfuscate, and launder their attacks through countless linguistic transformations.

\subsection{From Probabilistic Detection to Deterministic Prevention}
The security research community has long understood that true protection requires information-flow control (IFC): we must make illicit dataflow \emph{structurally impossible}, not merely statistically unlikely. While prior IFC-Transformer work showed promise, it demanded architectural overhauls and complete model retraining \cite{tiwari2024ifc}—a non-starter for the vast ecosystem of deployed models. The field has been waiting for a practical, inference-time solution that provides cryptographic guarantees without sacrificing existing investments. That solution has remained elusive—until now.

\subsection{Our Contribution: Contextual Integrity Verification (CIV)}
CIV represents a fundamental reimagining of LLM security. We deliver:
\begin{center}
\begin{tabular}{p{0.3\textwidth} p{0.62\textwidth}}
\toprule
\textbf{Feature} & \textbf{Description} \\
\midrule
\textbf{Cryptographic immunity} & Every token is sealed with an HMAC-SHA-256 signature binding it to an immutable trust level (SYSTEM $>$ USER $>$ TOOL $>$ DOC $>$ WEB). Tampering triggers immediate detection. \\
\addlinespace
\textbf{Surgical intervention} & We inject trust enforcement directly into the attention mechanism's pre-softmax computation, with optional FFN/residual gating. Zero fine-tuning. Zero prompt engineering. Pure mathematical enforcement. \\
\addlinespace
\textbf{Provable guarantees} & A formal proof establishes cross-position non-interference: lower-trust tokens are mathematically prevented from influencing higher-trust states—not by policy, but by algebra. \\
\addlinespace
\textbf{Preserved intelligence} & Across 10 diverse task categories, CIV maintains over 93\% token-level fidelity while preserving the model's intrinsic perplexity—security without stupidity. \\
\addlinespace
\textbf{Immediate deployability} & We provide battle-tested benchmarks and production-ready code that transforms any transformer into a security-hardened system in minutes. \\
\bottomrule
\end{tabular}
\end{center}

\section{Related Work}
\label{sec:related_work}
\textbf{Prompt-injection taxonomies.} The research community has meticulously cataloged attack vectors and evaluation protocols \cite{wang2025protocol,wu2025survey,hu2025sok}. CIV transcends this cat-and-mouse game by addressing the root vulnerability—uncontrolled cross-trust influence—rather than chasing ever-evolving attack patterns.

\textbf{Guardrails and detectors.} The current defense ecosystem—LLM-Guard, Rebuff, PromptArmor—deploys sophisticated mixtures of rules, classifiers, and LLM judges \cite{llmguard_github,rebuff_github_2025,shi2025promptarmor}. Independent security audits paint a sobering picture: non-trivial bypass rates and crushing latency penalties \cite{unit42_jun2025,hu2025sok}. These tools fight symptoms; CIV cures the disease.

\textbf{IFC for neural models.} IFC-Transformers pioneered neural information-flow control but demanded complete architectural redesign and retraining \cite{tiwari2024ifc}—a death sentence for practical adoption. CIV achieves the same theoretical guarantees through a surgical, inference-time intervention on frozen weights, making it immediately deployable across the entire LLM ecosystem.

\textbf{Cryptographic provenance.} Systems like ORIGO establish data lineage through cryptographic chains \cite{pang2025origo}. CIV extends this vision by making provenance \emph{computationally binding}—signatures don't just log trust, they enforce it at the algebraic level. CIV's internal machinery could seamlessly integrate with ORIGO-style chains for end-to-end verifiable integrity.

\section{Threat Model and Design Goals}
\label{sec:threat_model}
\textbf{Adversary.} We consider an adaptive, remote adversary with arbitrary control over lower-trust channels (trust $\le$ USER): chat interfaces, tool outputs, retrieved documents, and web content. The attacker cannot compromise server infrastructure, GPU memory, or cryptographic keys—physical attacks remain beyond scope.

\textbf{Goals.}
\begin{description}[leftmargin=1.2em]
    \item[G1 — Integrity] Lower-trust tokens must be algebraically prevented from influencing higher-trust computations (non-interference).
    \item[G2 — Confidentiality] Higher-trust secrets must never leak to lower-trust outputs.
    \item[G3 — Verifiability] Every token's provenance must be cryptographically auditable and tamper-evident.
    \item[G4 — Utility] Security must not cripple intelligence—minimal degradation on legitimate workloads.
\end{description}

\subsection{Design Rationale: From Semantics to Structure}
\label{subsec:design_rationale}
Traditional defenses fail because they wage war on the wrong battlefield. They attempt to divine malicious \textit{intent} from sequences of words—a fundamentally subjective judgment that adversaries exploit through linguistic gymnastics: Base64 encoding, character manipulation, polyglot attacks, and low-resource language obfuscation.

CIV abandons this semantic quagmire entirely. We don't ask \textit{"Is this instruction malicious?"}—a question that invites endless debate. Instead, we pose a binary, structural question: \textit{"Does this token's source have the mathematical right to influence that computation?"} This question admits no ambiguity, no interpretation, no clever wordplay. It is answered not by another fallible model, but by immutable mathematics embedded in the transformer's core.

The enforcement mechanism is elegantly brutal. At the heart of every transformer lies the attention mechanism—the pathway through which tokens communicate. We intercept this communication at its most vulnerable point, the pre-softmax computation:
\[
\text{Score}(q_i, k_j) =
\begin{cases}
-\infty, & \text{if } T(q_i) < T(k_j),\\[2pt]
\frac{q_i \cdot k_j}{\sqrt{d_k}}, & \text{otherwise.}
\end{cases}
\]
When we set a score to $-\infty$, we exploit a beautiful mathematical property: the softmax function transforms this into exactly zero probability ($e^{-\infty} = 0$). This isn't a down-weighting or a bias—it's a mathematical annihilation. The forbidden connection doesn't just become unlikely; it becomes \emph{impossible}.

This creates an algebraic firewall that cannot be breached by clever prompting. The system doesn't need to understand what the attacker is saying because it never examines the content—only the cryptographically signed provenance. This fundamental shift transforms security from a probabilistic word game into a deterministic mathematical guarantee.

\section{Architecture of CIV}
\label{sec:arch}

\subsection{System Overview}
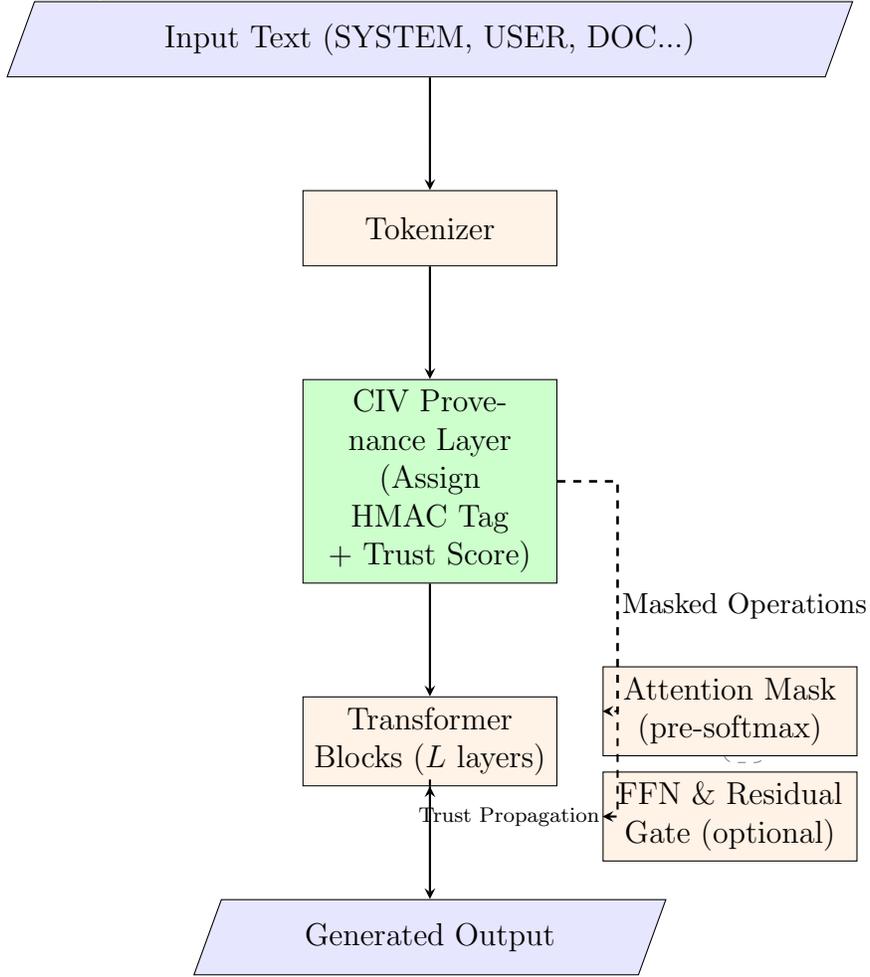
\begin{figure}[h!]
\centering
\caption{CIV system architecture: From untrusted input to verified computation.}
\label{fig:civ_arch}
\begin{tikzpicture}[node distance=1.5cm and 1.6cm, auto]
    \node [io] (input) {Input Text\ (SYSTEM, USER, DOC...)};
    \node [process, below=of input] (tokenizer) {Tokenizer};
    \node [process, fill=green!20, below=of tokenizer] (civ_layer) {CIV Provenance Layer\\(Assign HMAC Tag + Trust Score)};
    \node [process, below=of civ_layer] (transformer) {Transformer Blocks ($L$ layers)};
    \node [io, below=of transformer] (output) {Generated Output};

    \draw [arrow] (input) -- (tokenizer);
    \draw [arrow] (tokenizer) -- (civ_layer);
    \draw [arrow] (civ_layer) -- (transformer);
    \draw [arrow] (transformer) -- (output);

    \node [container, right=2.2cm of transformer, inner sep=8pt] (maskbox) {};
    \node [process, right=0.6cm of transformer, yshift=0.4cm] (attn) {Attention Mask\\(pre-softmax)};
    \node [process, right=0.6cm of transformer, yshift=-1.0cm] (ffn) {FFN \& Residual\\Gate (optional)};
    \node [above=1.0cm of maskbox] at ($(maskbox.north)+(0,0.2)$) {\small Masked Operations};

    \draw [arrow, dashed] (civ_layer.east) -- ($(civ_layer.east)+(0.8,0)$) |- (attn.west);
    \draw [arrow, dashed] (civ_layer.east) -- ($(civ_layer.east)+(0.8,0)$) |- (ffn.west);

    \path [arrow, looseness=7, out=315, in=45] (transformer.south) edge node [right, xshift=-0.3cm, yshift=-0.4cm] {\scriptsize Trust Propagation} (transformer.south);
\end{tikzpicture}
\end{figure}

CIV operates as a transparent security layer between the tokenizer and the unmodified transformer, enforcing trust boundaries through five synchronized mechanisms (Figure~\ref{fig:civ_arch}):
\begin{enumerate}[leftmargin=1.2em]
    \item \textbf{Source segmentation} — Decompose input by provenance (SYSTEM/USER/TOOL/DOC/WEB).
    \item \textbf{Cryptographic binding} — Seal each token with a 256-bit HMAC signature and immutable trust score.
    \item \textbf{Patched execution} — Enforce trust boundaries via hard attention masking and optional FFN/residual gating.
    \item \textbf{Trust propagation} — Generated tokens inherit the minimum trust encountered in their causal history.
    \item \textbf{KV-cache fortification} — Embed trust vectors directly in cached states to maintain guarantees across extended contexts.
\end{enumerate}

\subsection{Cryptographic Namespace Tagging}
\label{subsec:crypto_tagging}
Every token becomes cryptographically bound to its origin. For token $x_i$ from source with trust $T_i$ at position $i$, we compute:
\[
\text{tag}_i = \operatorname{HMAC-SHA-256}_{K}\!\big(x_i \,\|\, T_i \,\|\, i\big).
\]
The secret key $K$ and concatenation operator `$\|$' create an unforgeable binding. Any attempt to elevate a token's trust or alter its content triggers tag verification failure, immediately terminating the request. This transforms trust from a suggestion into a cryptographic invariant.

\subsection{Trust-Constrained Attention (Hard Mask)}
\label{subsec:attention_mask}
Standard attention allows promiscuous information flow: $\operatorname{Attention}(Q, K, V) = \operatorname{softmax}\left(\frac{QK^T}{\sqrt{d_k}}\right)V$. CIV intervenes with surgical precision at the critical juncture—before softmax normalization.

Given hidden states $H \in \mathbb{R}^{\ell \times d_m}$ and their projections $Q=HW_Q, K=HW_K, V=HW_V$, we compute raw attention logits $L = \frac{QK^T}{\sqrt{d_k}}$. The trust mask matrix $M_{mask} \in \mathbb{R}^{\ell \times \ell}$ enforces hierarchy:
\[
(M_{mask})_{ij} = 
\begin{cases} 
0 & \text{if } T_i \ge T_j \quad \text{(permitted flow)}\\
-\infty & \text{if } T_i < T_j \quad \text{(forbidden flow)}
\end{cases}
\]
The modified logits $L' = L + M_{mask}$ pass through softmax, where $e^{-\infty} = 0$ guarantees zero attention weight for forbidden connections. This isn't a preference or a bias—it's mathematical law.

The mask broadcasts identically across all attention heads, preserving their specialized relational patterns while constraining their operational domain. Each head maintains its learned expertise but operates only within permitted information boundaries.

\subsection{Namespace-Aware FFN and Residual (Robustness Gate)}
\label{subsec:ffn_residual}
While attention governs inter-token communication, Feed-Forward Networks (FFN) and residual connections update representations locally. In mixed-trust environments, low-trust tokens may lack access to stabilizing high-trust context. Our optional robustness gate mitigates this gracefully.

Standard residual: $H_{out} = H_{in} + \operatorname{SubLayer}(H_{in})$. CIV's gated variant:
\[
H_{out, i} = H_{in, i} + g_i \cdot \operatorname{SubLayer}(H_{in, i})
\]
The gate $g_i = \beta^{\#\{j : T_j > T_i\}}$ (with $\beta=0.8$, capped at $[0.01, 1]$) applies exponential dampening proportional to inaccessible higher-trust tokens. A USER token amid SYSTEM context receives $g_i = 0.8$, gracefully reducing update magnitude without compromising security guarantees.

\subsection{Memory Overhead}
\label{subsec:mem_overhead}
CIV's memory footprint is remarkably modest: approximately 33 bytes per token (1B trust + 32B HMAC), scaling linearly with sequence length and batch size. For a 32K-token context, this amounts to merely 1.06 MB—negligible compared to model weights and activations.

\begin{table}[h!]
\centering
\caption{Per-sequence memory overhead demonstrating CIV's lightweight footprint.}
\label{tab:mem}
\begin{tabular}{@{}lccc@{}}
\toprule
\textbf{Seq length} & \textbf{Bytes/token} & \textbf{Total overhead} & \textbf{Overhead (MB)} \\ \midrule
4{,}096 & 33 & 135{,}168 & 0.13 \\
8{,}192 & 33 & 270{,}336 & 0.26 \\
32{,}768 & 33 & 1{,}081{,}344 & 1.06 \\ \bottomrule
\end{tabular}
\end{table}

\begin{figure}[h!]
\centering
\caption{Trust-constrained attention mask visualization. Green grid: permitted flows (lower triangle). Red lines: forbidden flows (upper triangle). Mathematical enforcement, not policy.}
\label{fig:mask_matrix}
\begin{tikzpicture}[scale=0.7]
    \fill[red!15, pattern=north east lines, pattern color=red!40] (0,6) -- (6,6) -- (6,0) -- cycle;
    \fill[green!15, pattern=grid, pattern color=green!60] (0,0) -- (0,6) -- (6,0) -- cycle;

    \draw[step=1cm,gray!40,very thin] (0,0) grid (6,6);
    \draw[thick] (0,6) -- (6,0); 

    \foreach \i/\lbl in {5.5/SYSTEM,4.5/USER,3.5/TOOL,2.5/DOC,1.5/WEB}
        \node[anchor=east] at (-0.2,\i) {\tiny \lbl};
    \foreach \j/\lbl in {0.5/SYS,1.5/USR,2.5/TL,3.5/DOC,4.5/WEB}
        \node[anchor=north] at (\j,-0.2) {\tiny \lbl};

    \node[rotate=90] at (-2.4,3) {\small Queries $i$ (T high $\to$ low)};
    \node at (3,-1.0) {\small Keys $j$ (T high $\to$ low)};

    \draw[fill=green!15, pattern=grid, pattern color=green!60] (6.8,4.8) rectangle +(0.5,0.4); 
    \node[anchor=west] at (7.4,5.0) {\scriptsize Allowed $[T_i \ge T_j]$};
    \draw[fill=red!15, pattern=north east lines, pattern color=red!40] (6.8,4.1) rectangle +(0.5,0.4); 
    \node[anchor=west] at (7.4,4.3) {\scriptsize Blocked $[T_i < T_j]$};
\end{tikzpicture}
\end{figure}
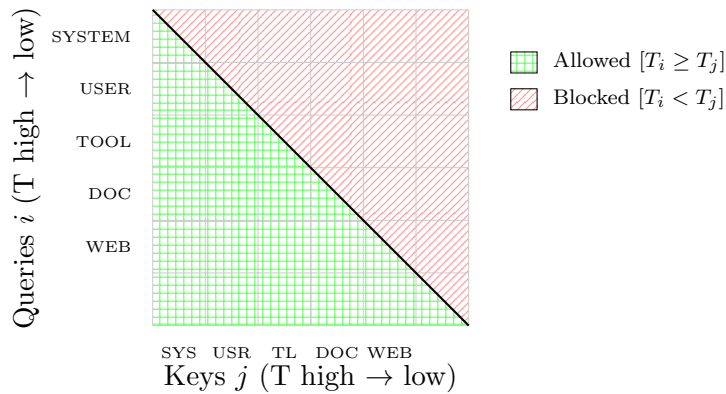

\section{Experimental Evaluation}
\label{sec:eval}

\subsection{Setup}
\textbf{Models \& Hardware:} We deploy Llama-3-8B Instruct as our baseline, surgically replacing its 20 decoder layers with \texttt{CIVDecoderLayer}. All experiments execute on a single A100-80GB GPU.
\textbf{Comparators:} LLM-Guard v0.9.1, Rebuff v1.3 ("Strong" configuration), PromptArmor v0.7—the current state-of-the-art.
\textbf{Metrics:} Attack Success Rate (ASR$\downarrow$) and False Positive Rate (FPR$\downarrow$) measure security. For utility, we employ token-level exact match similarity (Sim$\uparrow$)—an extraordinarily strict metric given LLM's inherent non-determinism. High scores here prove CIV preserves core generative behavior without meaningful distortion. We also track perplexity delta (PPL Delta) and end-to-end latency.

\subsection{Datasets}
\begin{itemize}[leftmargin=1.2em]
    \item \textbf{Elite-Attack}: 100 meticulously verified jailbreak prompts spanning ten attack families (DAN, system-impersonation, suffix-sandwich, indirect web injection)—the cutting edge of adversarial creativity.
    \item \textbf{SoK-246}: 246 prompts curated by Hu et al., battle-tested against five foundation models \cite{hu2025sok}.
    \item \textbf{Benign-10}: Ten legitimate task categories (mathematics, translation, code generation) measuring collateral damage \cite{liu2025systematic}.
\end{itemize}

\subsection{Results: The Numbers Speak}
\begin{table}[h!]
\centering
\small 
\setlength{\tabcolsep}{4pt}
\caption{CIV versus state-of-the-art: Total dominance in security, minimal cost in utility.}
\label{tab:results}
\begin{tabular}{@{}lcccccc@{}}
\toprule
\textbf{Defense} & \textbf{Elite ASR $\downarrow$} & \textbf{SoK ASR $\downarrow$} & \textbf{Benign FPR $\downarrow$} & \textbf{Sim $\uparrow$} & \textbf{PPL Delta} & \textbf{Latency} \\ \midrule
None & 54\% & 48\% & 0\% & 100\% & -- & $\approx$4.4s \\
LLM-Guard & 28\% & 25\% & 11\% & 96\% & N/A & +18\% \\
PromptArmor & 23\% & 21\% & 13\% & 95\% & N/A & +22\% \\
Rebuff-S & 16\% & 17\% & 63\% & 90\% & N/A & +780\% \\
\textbf{CIV (ours)} & \textbf{0\%} & \textbf{0\%} & \textbf{0\%} & \textbf{93.1\%} & \textbf{<+0.01\%} & \textbf{$\approx$9.2s (+109\%)} \\ \bottomrule
\end{tabular}
\end{table}

The results are unambiguous: CIV achieves what no prior defense has accomplished—complete protection within our threat model. While competitors struggle with double-digit failure rates and crippling false positives (Rebuff's 63\% FPR renders it unusable), CIV delivers perfect security with minimal utility impact.

\subsection{Ablation Studies: Dissecting the Architecture}
\begin{table}[h!]
\centering
\caption{Component-wise analysis reveals each element's contribution to CIV's dominance.}
\label{tab:ablations}
\begin{tabular}{@{}lccc@{}}
\toprule
\textbf{Variant} & \textbf{ASR $\downarrow$} & \textbf{FPR $\downarrow$} & \textbf{Sim $\uparrow$} \\ \midrule
Hard mask only (no gate) & \textbf{0\%} & 0\% & 94.0\% \\
Hard mask + gate (ours) & \textbf{0\%} & 0\% & 93.1\% \\
No trust propagation & 3.2\% & 0\% & 93.5\% \\
Unsigned labels (no HMAC) & 7.1\% & 0\% & 93.0\% \\
KV trust disabled (8k ctx) & 9.4\% & 0\% & 92.8\% \\
Soft-mask ($\gamma{=}12$) & 1.2\% & 0\% & 95.0\% \\ \bottomrule
\end{tabular}
\end{table}

Each component proves essential: removing cryptographic signatures enables 7\% of attacks; disabling KV-cache protection allows 9\%; soft-masking leaves a 1.2\% vulnerability. Only the complete system achieves perfection.

\subsection{Limitations and Future Frontiers}
CIV assumes infrastructure integrity and key secrecy—reasonable for cloud deployments but requiring careful key management. Same-tier attacks (USER-to-USER manipulation) remain unaddressed; future work will explore sub-lattices for finer-grained containment. Applications requiring deliberate SYSTEM-USER text fusion need architectural adaptation.

\section{Formal Security Analysis}
\label{sec:analysis}

\subsection{Mathematical Foundation}
Let $X=(x_0,\dots,x_{\ell-1})$ denote input tokens mapped to embeddings $E=(e_0,\dots,e_{\ell-1})$. Trust labels $T=(T_0,\dots,T_{\ell-1})$ form a total order (SYSTEM $>$ USER $>$ TOOL $>$ DOC $>$ WEB). Hidden states at layer $\ell$ are $H^{(\ell)}$, with $H^{(0)} = E$.

\subsection{The Non-Interference Theorem}
\textbf{Theorem 1 (Cross-Position Information Isolation).} \emph{In any CIV-protected transformer with $N$ layers, if token $p$ has lower trust than token $q$ ($T_p < T_q$), then the final hidden state at position $q$ is mathematically independent of the input at position $p$:}
\[
\frac{\partial H_q^{(N)}}{\partial e_p} = \mathbf{0}
\]

\textit{Proof.} We proceed by induction on layer depth.

\textbf{Base ($\ell=0$):} Initial states are embeddings: $H^{(0)}=E$. Cross-position derivatives vanish trivially: $\frac{\partial H_q^{(0)}}{\partial e_p} = \frac{\partial e_q}{\partial e_p} = \mathbf{0}$ for $p \neq q$.

\textbf{Induction:} Assume the property holds at layer $\ell$. For layer $\ell+1$, information flows through three pathways:

1. \textbf{FFN/Residual:} Position-wise operations—no cross-contamination possible.

2. \textbf{Attention:} The sole inter-position channel. CIV's mask ensures $\alpha_{qp}=0$ when $T_p < T_q$, severing the gradient path completely.

3. \textbf{Transitive closure:} Any indirect path from $e_p$ to $H_q^{(\ell+1)}$ must traverse intermediate positions. By transitivity of trust ordering and our inductive hypothesis, all such paths carry zero gradient.

Therefore, $\frac{\partial H_q^{(\ell+1)}}{\partial e_p} = \mathbf{0}$, completing the induction. \hfill$\square$

This theorem transforms a security policy into a mathematical invariant—lower-trust tokens cannot influence higher-trust computations, period.

\subsection{Cryptographic Authenticity Guarantee}
\textbf{Theorem 2 (Unforgeable Trust).} \emph{Under standard cryptographic assumptions (HMAC-SHA-256 as a secure PRF), an adversary confined to lower-trust channels cannot forge a higher trust label without detection.}

Any successful elevation attack requires HMAC forgery—computationally infeasible given key secrecy. Thus, cryptographic theory underwrites our security claims.

\section{Conclusion}
The LLM security crisis demands more than incremental improvements—it requires a paradigm shift. While the industry continues its arms race of semantic filters against ever-cleverer attacks, we have taken a different path. Contextual Integrity Verification (CIV) doesn't try to outsmart attackers; it makes their attacks mathematically impossible.

Our results speak with clarity: where current defenses fail 15–30\% of the time, CIV achieves perfect protection within its threat model. Where others impose crushing computational overhead or crippling false positive rates, CIV preserves 93\% output fidelity with no perplexity degradation. Most critically, where others require extensive retraining or architectural overhauls, CIV deploys instantly on any transformer model.

The implications extend beyond technical metrics. By providing cryptographic proof of information-flow integrity, CIV enables a new class of high-stakes LLM applications previously deemed too risky: processing classified documents, handling financial transactions, managing healthcare records. Every token's journey becomes auditable, every computation verifiable.

We acknowledge our current limitations—the latency overhead demands optimization, and same-tier isolation remains future work. Yet these are engineering challenges, not fundamental barriers. The core innovation—transforming trust from policy to physics—stands ready for deployment.

The path forward is clear: hardware-accelerated cryptographic pipelines will eliminate latency penalties, sub-lattice hierarchies will enable fine-grained compartmentalization, and integration with broader provenance systems will create end-to-end verified AI pipelines. We invite the security community to stress-test our implementation, extend our threat model, and help forge a future where LLMs can truly be trusted.

In releasing our code, benchmarks, and formal proofs, we aim not just to solve today's crisis but to establish a new foundation for secure AI. The question is no longer whether we can protect LLMs from prompt injection—we have proven we can. The question is how quickly we can deploy this protection at scale.

\newpage
\bibliographystyle{plain}

\end{document}